# Microcavities integrated in metal halide perovskite light-emitting field-effect transistors

Francesco Scotognella

Dipartimento di Fisica, Politecnico di Milano, Piazza Leonardo da Vinci 32, 20133 Milano, Italy
e-mail address: francesco.scotognella@polimi.it

**Abstract**

Metal halide perovskites are materials that show unique characteristics for photovoltaics and light emission. Amplified spontaneous emission and stimulated emission has been shown with these materials, together with electroluminescence in light-emitting diodes and light-emitting transistors. An important achievement that combine stimulated emission and electroluminescence could be the fabrication of electrically driven metal halide perovskite lasers.

In this work, the integration of metal halide perovskite light-emitting field-effect transistors with photonic microcavities is proposed. This can lead to the engineering of electrically driven lasers. The microcavities have been designed in order to have the cavity mode at 750 nm, which is the peak wavelength of the electroluminescent spectrum of recently reported $MaPbI_3$-based electroluminescent devices. The optical properties of the photonic microcavities have been simulated by means of the transfer matrix method, considering the wavelength dependent refractive indexes of all the materials involved. The material for the gate is indium tin oxide, while different materials, either inorganic or organic, have been considered for the microcavity architectures.

**Introduction**

Metal halide perovskites show ground-breaking properties in terms of photovoltaic efficiency and light emission [1]. In April 2021, a power conversion efficiency of 25.6% has been reached with a perovskite solar cell [2]. In July 2022, 31.25% efficiency for tandem perovskite-silicon solar cell has been achieved [3]. Concerning light emission, metal halide perovskite based light emitting diodes are very established [4,5]. Furthermore, amplified spontaneous emission and laser emission has been observed in many reports with metal halide perovskites [6–8]. The combination of laser emission and electroluminescence is very challenging, but it can lead to the fabrication of electrically driven perovskite lasers, a breakthrough for many applications from photonics to communications. Interesting structures are proposed to achieve such device [9].

The use of light-emitting field-effect transistors could be strategic for electrically driven lasing, since they allow a high carrier density together with the control of current flow, charge injection and emission patterns [10,11]. The ambipolar characteristics of metal halide perovskites, such as methylammonium lead iodide (CH3NH3PbI3 or MAPbI3), allow the fabrication of MAPbI3-based light-emitting transistors [12–14]. The integration of a laser cavity in light-emitting diodes could be pursued via the inclusion of photonic crystals in the transistors. Organic light-emitting transistors with photonic crystal gate dielectrics have been proposed in previous reports [15,16], demonstrating electroluminescence spectral modulation and enhancement.

In this work, we have designed different microcavities integrated with light-emitting field effect transistors based on metal halide perovskites. These architectures could allow the fabrication of electrically driven metal halide perovskite lasers in which the microcavities provide the feedback mechanism. Different materials are employed to widen the fabrication possibilities: inorganic materials such as silicon dioxide, titanium dioxide, fluorine indium co-doped cadmium oxide (FICO), silicon; organic polymers as polyvinyl carbazole (PVK), cellulose acetate (CA). The transfer matrix method has been employed considering all the wavelength dependent refractive indexes of the materials.

**Methods**

In this work the transmission spectra of the microcavities have been simulated with the transfer matrix method, a well establish tool to study one dimensional photonic structures [17,18]. Within this method the matrix used for the $k$th layer is given by

$$M_k = \begin{bmatrix} \cos\left(\frac{2\pi}{\lambda}n_k(\lambda)d_k\right) & -\frac{i}{n_k(\lambda)}\sin\left(\frac{2\pi}{\lambda}n_k(\lambda)d_k\right) \\ -in_k(\lambda)\sin\left(\frac{2\pi}{\lambda}n_k(\lambda)d_k\right) & \cos\left(\frac{2\pi}{\lambda}n_k(\lambda)d_k\right) \end{bmatrix} \quad (1)$$

with $n_k(\lambda)$ the wavelength dependent refractive index of the layer and $d_k$ the thickness of the layer (in nm). The matrix describing the whole multilayer system is given by the product of the $M_k$ matrices

$$M = \prod_{i=1}^{N} M_k = \begin{bmatrix} M_{11} & M_{12} \\ M_{21} & M_{22} \end{bmatrix} \quad (2)$$

The integer number $N$ represents the number of layers. From the matrix elements of the matrix $M$ it is possible to compute the transmission coefficient

$$t = \frac{2n_s}{(M_{11}+M_{12}n_0)n_s+(M_{21}+M_{22}n_0)} \quad (3)$$

and, consequently, the transmission

$$T = \frac{n_0}{n_s}|t|^2 \quad (4)$$

In Equations 4 and 5, $n_0$ and $n_s$ are the refractive indexes of air and glass, respectively.
For silicon dioxide, the following Sellmeier equation has been employed [19,20]

$$n_{SiO_2}^2(\lambda) - 1 = \frac{0.6961663\lambda^2}{\lambda^2-0.0684043^2} + \frac{0.4079426\lambda^2}{\lambda^2-0.1162414^2} + \frac{0.8974794\lambda^2}{\lambda^2-9.896161^2} \quad (5)$$

For titanium dioxide, the wavelength-dependent refractive index is given by [21]

$$n_{TiO_2}(\lambda) = \left(4.99 + \frac{1}{96.6\lambda^{1.1}} + \frac{1}{4.60\lambda^{1.95}}\right)^{1/2} \quad (6)$$

For PVK, the Sellmeier equation for the refractive index of is given by [22,23]:

$$n_{PVK}^2(\lambda) - 1 = \frac{0.09788\lambda^2}{\lambda^2-0.3257^2} + \frac{0.6901\lambda^2}{\lambda^2-0.1419^2} + \frac{0.8513\lambda^2}{\lambda^2-1.1417^2} \quad (7)$$

For CA, the Sellmeier equation for the refractive index of CA is [22,23]:

$$n_{CA}^2(\lambda) - 1 = \frac{0.6481\lambda^2}{\lambda^2-0.0365^2} + \frac{0.5224\lambda^2}{\lambda^2-0.1367^2} + \frac{2.483\lambda^2}{\lambda^2-13.54^2} \quad (8)$$

In the equations 5 to 8 $\lambda$ is in micrometers.
To predict the behaviour of the plasmonic response in our photonic structures the Drude model can be employed [24,25], where the frequency dependent complex dielectric function of ITO and FICO can be written as:

$$\varepsilon_{FICO}(\omega) = \varepsilon_1(\omega) + i\varepsilon_2(\omega) \quad (9)$$

where

$$\varepsilon_1 = \varepsilon_\infty - \frac{\omega_P^2}{(\omega^2-\Gamma^2)} \quad (10)$$

and

$$\varepsilon_2 = \frac{\omega_P^2\Gamma}{\omega(\omega^2-\Gamma^2)} \quad (11)$$

with

$$\omega_P = \sqrt{\frac{Ne^2}{\varepsilon_0 m^*}} \qquad (12)$$

For ITO $N = 2.49 \times 10^{26} \, cm^{-3}$, $\varepsilon_\infty = 4$, $m^* = 0.4 m_e$ and $\Gamma = 0.1132 \, eV$ [26]. For FICO $N = 1.68 \times 10^{27} \, cm^{-3}$, $\varepsilon_\infty = 5.6$, $m^* = 0.43 m_e$ and $\Gamma = 0.07 \, eV$ [26,27].

The wavelength dependent refractive index of silicon is taken from Ref. [28]. Considering the studied wavelength range for the microcavity that includes silicon layers (500 – 1500 nm), the imaginary part of the refractive index is neglected. Finally, the wavelength dependent refractive of the active material MAPbI$_3$ is taken from Refs. [29,30]. Such refractive index dispersion includes the real part and the imaginary part.

**Results and Discussion**

The typical structure of a light-emitting transistor integrated with a microcavity is depicted in Figure 1, in agreement with the structures fabricated in Refs. [15,16]. As gate contact, indium tin oxide has been selected. The source and drain contacts are placed at the two sides of the MAPbI$_3$ layer. Several materials can be used as source and drain, such as gold/nickel [12] or gold/titanium [31].

The selected thickness for the MAPbI$_3$ layer is 40 nm, as reported in the fabrication of a MAPbI$_3$-based light-emitting diode [5]. To tune the thickness of the defect, the MAPbI$_3$ layer is embedded between two layers of another material. In the example in Figure 1, the two one dimensional photonic crystals are made of titanium dioxide and FICO, while the MAPbI$_3$ layer is embedded between two layers of silicon dioxide. The employment of FICO is interesting since it could allow also the possibility of UV light-induced dielectric function change via photodoping [27].

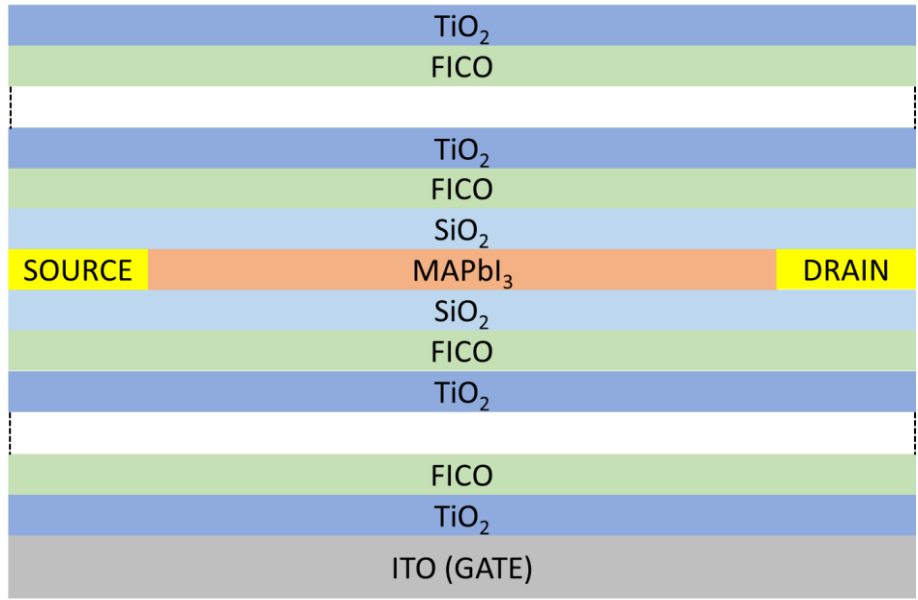

**Figure 1**. Structure of the MAPbI$_3$ light-emitting field-effect transistor integrated with a microcavity composed by two FICO/TiO$_2$ one-dimensional photonic crystals with the MAPbI$_3$ layer as defect.

In Figure 2 the transmission spectrum of the structure depicted in Figure 1 is shown. The one-dimensional photonic crystals have six bilayers of TiO$_2$ and FICO. The cavity mode with the highest transmission is centred at 750 nm, where previously reported MAPbI$_3$-based light-emitting diodes show the peak wavelength of the electroluminescent spectrum [5,9]. The microcavity shows an additional cavity mode at 652 nm, very weak in terms of transmission.

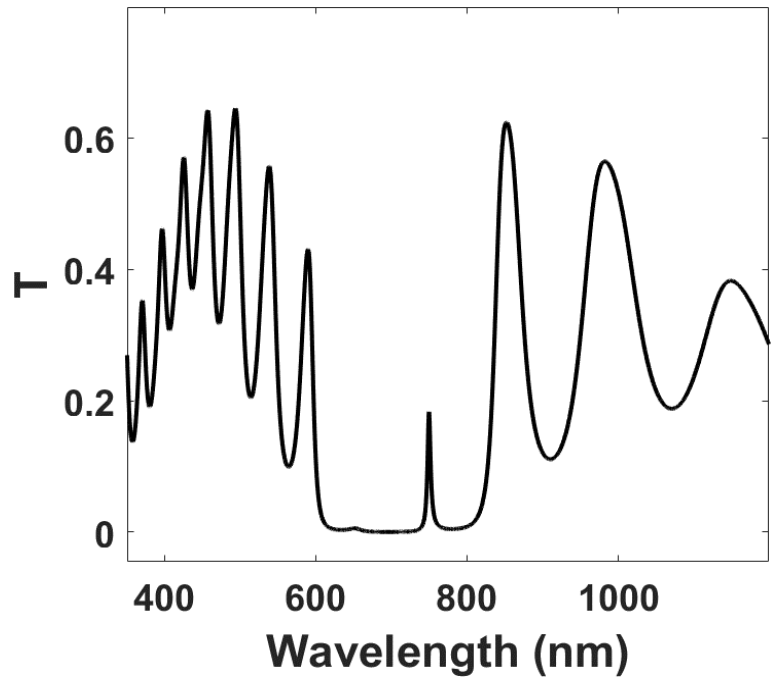

**Figure 2**. Transmission spectrum of ITO/(TiO$_2$/FICO)$_6$/SiO$_2$/MaPbI$_3$/SiO$_2$/(FICO/TiO2)$_6$ .

The thickness of the ITO layer is 100 nm. The thickness of the titanium dioxide layers is 73.5 nm, while the thickness of the FICO layers is 112.7 nm. With 6 bilayers of TiO$_2$ and FICO, the total thickness of each photonic crystal is 1117 nm. In order to have the cavity mode at 750 nm, the MaPbI$_3$ layer is embedded between two layers of silicon dioxide with a thickness of 144.2 nm. In this structure, the weaker defect mode at 652 nm is more evident.

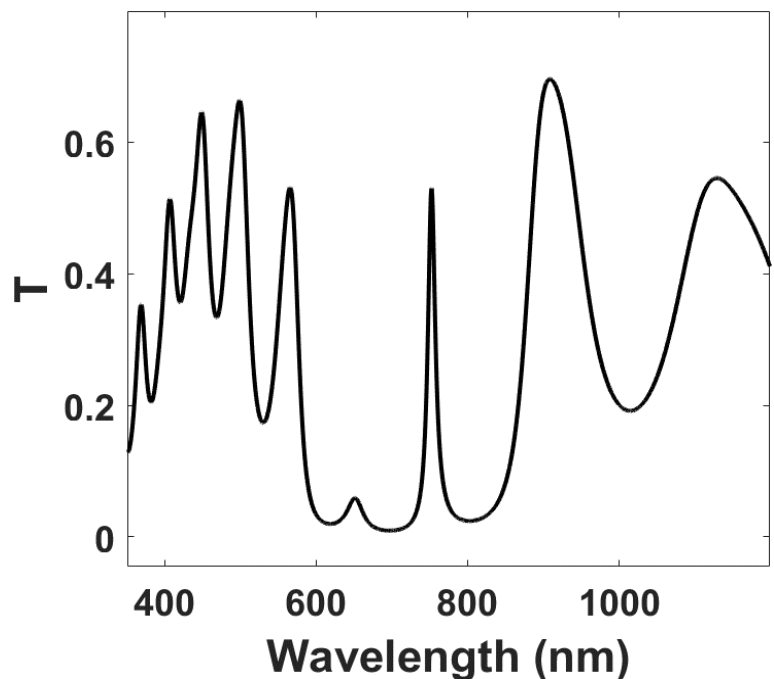

**Figure 3**. Transmission spectrum of ITO/(TiO$_2$/FICO)$_4$/SiO$_2$/MaPbI$_3$/SiO$_2$/(FICO/TiO2)$_4$ .

Since the photonic crystal between the ITO-based gate and the active material, i.e. MaPbI$_3$, has to act as a gate dielectric, its thickness is crucial for the operation of the light-emitting transistor. Moreover, also the thickness silicon dioxide buffer layer within the cavity should be considered. The thickness can be controlled with the number of bilayers in the photonic crystals. With 4 bilayers of TiO$_2$ and FICO, the total thickness of each photonic crystal is 745 nm, in line with the thickness employed in a previously reported light-emitting transistor [16].

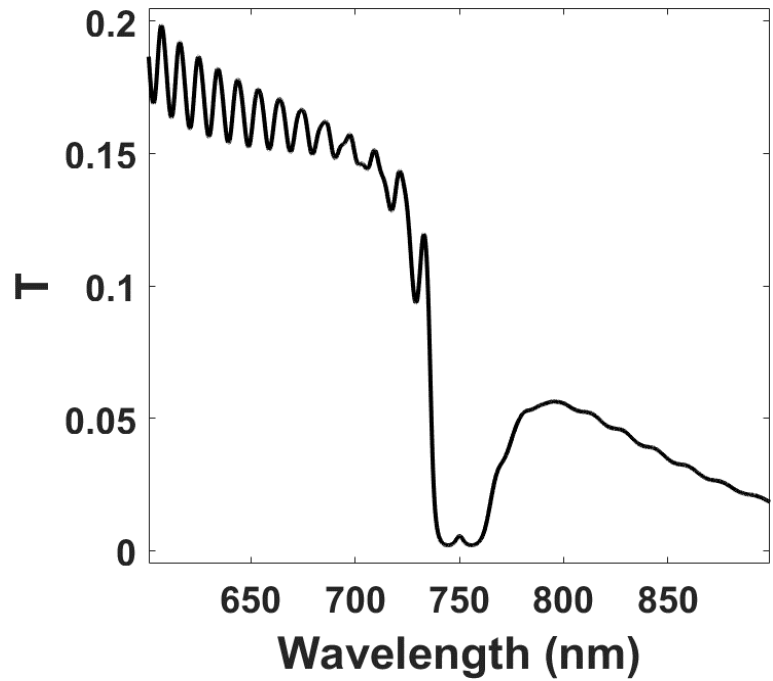

**Figure 4**. Transmission spectrum of ITO/(SiO$_2$/FICO)$_{50}$/TiO$_2$/MaPbI$_3$/TiO$_2$/(FICO/SiO$_2$)$_{50}$ .

A microcavity in which the one-dimensional photonic crystal is made of SiO$_2$ and FICO and the defect is made of TiO$_2$/MaPbI$_3$/TiO$_2$ shows a transmission spectrum as in Figure 4. The thickness of the silicon dioxide layers is 128.4 nm, while the thickness of FICO is 124.1 nm. The thickness of the titanium dioxide layers is 139.5 nm. Because of the similar real part of the refractive index of silicon dioxide and FICO, the photonic band gap is relatively narrow and a high number of bilayers is needed to achieve a sufficiently efficient photonic band gap. Because of the high number of layers of FICO the transmission strongly decreases in the near infrared.

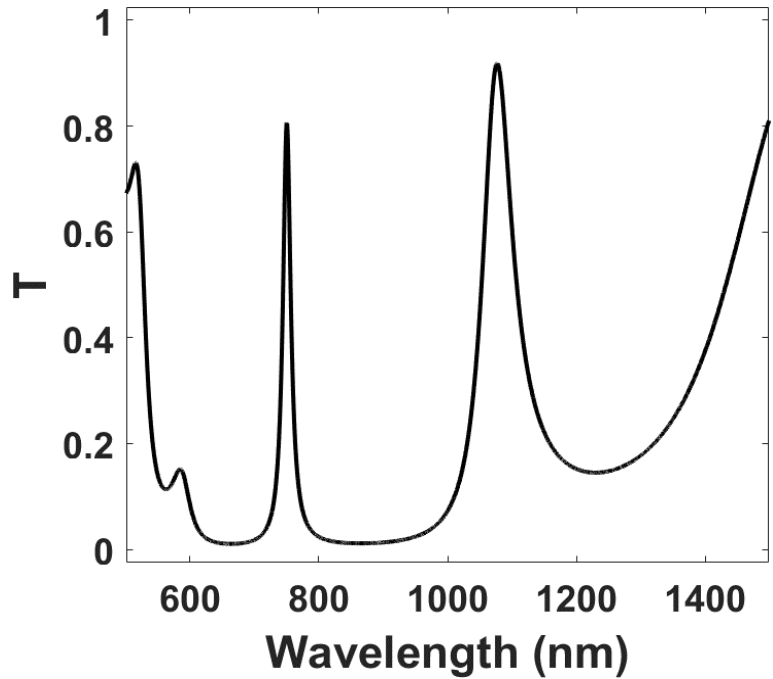

**Figure 5**. Transmission spectrum of ITO/(SiO$_2$/Si)$_2$/TiO$_2$/MaPbI$_3$/TiO$_2$/(Si/SiO$_2$)$_2$ .

To achieve very thin photonic crystals, and thus a very thin gate dielectric ITO and MaPbI$_3$, a structure alternating silicon dioxide and silicon is designed. The thickness of the silicon layers is 53.1 nm, while the thickness of the silicon dioxide layers is 123.9 nm. The thickness of the titanium dioxide layers in the microcavity defect is 139.5 nm. Thus, the total thickness of the dielectric between ITO and MaPbI$_3$ is 493.5 nm.

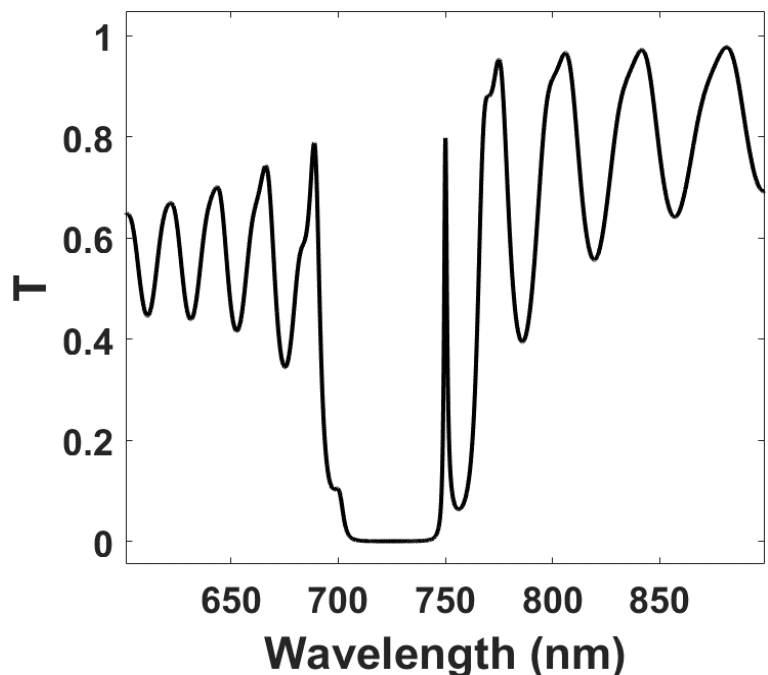

**Figure 6.** Transmission spectrum of ITO/(CA/PVK)$_{25}$/CA/MaPbI$_3$/CA/(PVK/CA)$_{25}$ .

Finally, a microcavity with fully organic photonic crystals has been designed, allowing completely different fabrication techniques such as spin coating [32]. The two one-dimensional photonic crystals are made of PVK and CA. With these two materials, the refractive index is quite high for polymers. However, to achieve an efficient photonic band gap, a high number of bilayers is needed.

**Conclusion**

Different microcavities integrated with metal halide based light-emitting field-effect transistors have been designed by means of transfer matrix method. Different materials have been employed in order to provide a library of diverse architectures. The architectures include ITO as gate dielectric and the active material MaPbI$_3$ is embedded between two one-dimensional photonic crystals. The photonic crystals have been made with silicon, silicon dioxide, titanium dioxide, FICO, PVK and CA. With the couple Si/SiO$_2$, very thin photonic crystals have been engineered. These architectures could allow the fabrication of electrically driven metal halide perovskite lasers.

**References**

[1] J. Sun, J. Wu, X. Tong, F. Lin, Y. Wang, Z.M. Wang, Organic/Inorganic Metal Halide Perovskite Optoelectronic Devices beyond Solar Cells, Advanced Science. 5 (2018) 1700780. https://doi.org/10.1002/advs.201700780.

[2] J. Jeong, M. Kim, J. Seo, H. Lu, P. Ahlawat, A. Mishra, Y. Yang, M.A. Hope, F.T. Eickemeyer, M. Kim, Y.J. Yoon, I.W. Choi, B.P. Darwich, S.J. Choi, Y. Jo, J.H. Lee, B. Walker, S.M. Zakeeruddin, L. Emsley, U. Rothlisberger, A. Hagfeldt, D.S. Kim, M. Grätzel, J.Y. Kim, Pseudo-halide anion engineering for α-FAPbI3 perovskite solar cells, Nature. 592 (2021) 381–385. https://doi.org/10.1038/s41586-021-03406-5.

[3] E. Bellini, CSEM, EPFL achieve 31.25% efficiency for tandem perovskite-silicon solar cell, Pv Magazine International. (n.d.). https://www.pv-magazine.com/2022/07/07/csem-epfl-achieve-31-25-efficiency-for-tandem-perovskite-silicon-solar-cell/ (accessed July 26, 2022).

[4] N.K. Kumawat, D. Gupta, D. Kabra, Recent Advances in Metal Halide-Based Perovskite Light-Emitting Diodes, Energy Technology. 5 (2017) 1734–1749. https://doi.org/10.1002/ente.201700356.

[5] L. Zhao, K. Roh, S. Kacmoli, K. Al Kurdi, S. Jhulki, S. Barlow, S.R. Marder, C. Gmachl, B.P. Rand, Thermal Management Enables Bright and Stable Perovskite Light-Emitting Diodes, Advanced Materials. 32 (2020) 2000752. https://doi.org/10.1002/adma.202000752.

[6] F. Deschler, M. Price, S. Pathak, L.E. Klintberg, D.-D. Jarausch, R. Higler, S. Hüttner, T. Leijtens, S.D. Stranks, H.J. Snaith, M. Atatüre, R.T. Phillips, R.H. Friend, High Photoluminescence Efficiency and Optically Pumped Lasing in Solution-Processed Mixed Halide Perovskite Semiconductors, J. Phys. Chem. Lett. 5 (2014) 1421–1426. https://doi.org/10.1021/jz5005285.

[7] P.J. Cegielski, A.L. Giesecke, S. Neutzner, C. Porschatis, M. Gandini, D. Schall, C.A.R. Perini, J. Bolten, S. Suckow, S. Kataria, B. Chmielak, T. Wahlbrink, A. Petrozza, M.C. Lemme, Monolithically Integrated Perovskite Semiconductor Lasers on Silicon Photonic Chips by Scalable Top-Down Fabrication, Nano Lett. 18 (2018) 6915–6923. https://doi.org/10.1021/acs.nanolett.8b02811.

[8] A.L. Alvarado-Leaños, D. Cortecchia, G. Folpini, A.R.S. Kandada, A. Petrozza, Optical Gain of Lead Halide Perovskites Measured via the Variable Stripe Length Method: What We Can Learn and How to Avoid Pitfalls, Advanced Optical Materials. n/a (2021) 2001773. https://doi.org/10.1002/adom.202001773.

[9] W. Gao, S.F. Yu, Reality or fantasy—Perovskite semiconductor laser diodes, EcoMat. 3 (2021) e12077. https://doi.org/10.1002/eom2.12077.

[10] M. Prosa, S. Moschetto, E. Benvenuti, M. Zambianchi, M. Muccini, M. Melucci, S. Toffanin, 2,3-Thienoimide-ended oligothiophenes as ambipolar semiconductors for multifunctional

single-layer light-emitting transistors, Journal of Materials Chemistry C. 8 (2020) 15048–15066. https://doi.org/10.1039/D0TC03326J.

[11]     H. Chen, W. Huang, T.J. Marks, A. Facchetti, H. Meng, Recent Advances in Multi-Layer Light-Emitting Heterostructure Transistors, Small. 17 (2021) 2007661. https://doi.org/10.1002/smll.202007661.

[12]     X.Y. Chin, D. Cortecchia, J. Yin, A. Bruno, C. Soci, Lead iodide perovskite light-emitting field-effect transistor, Nat Commun. 6 (2015) 7383. https://doi.org/10.1038/ncomms8383.

[13]     F. Maddalena, X.Y. Chin, D. Cortecchia, A. Bruno, C. Soci, Brightness Enhancement in Pulsed-Operated Perovskite Light-Emitting Transistors, ACS Appl. Mater. Interfaces. 10 (2018) 37316–37325. https://doi.org/10.1021/acsami.8b11057.

[14]     M. Klein, J. Li, A. Bruno, C. Soci, Co-Evaporated Perovskite Light-Emitting Transistor Operating at Room Temperature, Advanced Electronic Materials. 7 (2021) 2100403. https://doi.org/10.1002/aelm.202100403.

[15]     E.B. Namdas, B.B.Y. Hsu, J.D. Yuen, I.D.W. Samuel, A.J. Heeger, Optoelectronic Gate Dielectrics for High Brightness and High-Efficiency Light-Emitting Transistors, Advanced Materials. 23 (2011) 2353–2356. https://doi.org/10.1002/adma.201004102.

[16]     M. Natali, S.D. Quiroga, L. Passoni, L. Criante, E. Benvenuti, G. Bolognini, L. Favaretto, M. Melucci, M. Muccini, F. Scotognella, F.D. Fonzo, S. Toffanin, Simultaneous Tenfold Brightness Enhancement and Emitted-Light Spectral Tunability in Transparent Ambipolar Organic Light-Emitting Transistor by Integration of High-k Photonic Crystal, Advanced Functional Materials. 27 (2017) 1605164. https://doi.org/10.1002/adfm.201605164.

[17]     M. Born, E. Wolf, A.B. Bhatia, P.C. Clemmow, D. Gabor, A.R. Stokes, A.M. Taylor, P.A. Wayman, W.L. Wilcock, Principles of Optics: Electromagnetic Theory of Propagation, Interference and Diffraction of Light, 7th ed., Cambridge University Press, 1999. https://doi.org/10.1017/CBO9781139644181.

[18]     M. Bellingeri, A. Chiasera, I. Kriegel, F. Scotognella, Optical properties of periodic, quasi-periodic, and disordered one-dimensional photonic structures, Optical Materials. 72 (2017) 403–421. https://doi.org/10.1016/j.optmat.2017.06.033.

[19]     I.H. Malitson, Interspecimen Comparison of the Refractive Index of Fused Silica*,†, J. Opt. Soc. Am., JOSA. 55 (1965) 1205–1209. https://doi.org/10.1364/JOSA.55.001205.

[20]     RefractiveIndex.INFO - Refractive index database, (n.d.). https://refractiveindex.info/ (accessed November 15, 2019).

[21]     F. Scotognella, A. Chiasera, L. Criante, E. Aluicio-Sarduy, S. Varas, S. Pelli, A. Łukowiak, G.C. Righini, R. Ramponi, M. Ferrari, Metal oxide one dimensional photonic crystals made by RF sputtering and spin coating, Ceramics International. 41 (2015) 8655–8659. https://doi.org/10.1016/j.ceramint.2015.03.077.

[22]     L. Fornasari, F. Floris, M. Patrini, D. Comoretto, F. Marabelli, Demonstration of fluorescence enhancement via Bloch surface waves in all-polymer multilayer structures, Phys. Chem. Chem. Phys. 18 (2016) 14086–14093. https://doi.org/10.1039/C5CP07660A.

[23]     G. Manfredi, C. Mayrhofer, G. Kothleitner, R. Schennach, D. Comoretto, Cellulose ternary photonic crystal created by solution processing, Cellulose. 23 (2016) 2853–2862. https://doi.org/10.1007/s10570-016-1031-x.

[24]     J. Müller, C. Sönnichsen, H. von Poschinger, G. von Plessen, T.A. Klar, J. Feldmann, Electrically controlled light scattering with single metal nanoparticles, Applied Physics Letters. 81 (2002) 171. https://doi.org/10.1063/1.1491003.

[25]     L. Novotny, B. Hecht, Principles of nano-optics, 2. ed, Cambridge Univ. Press, Cambridge, 2012.

[26] I. Kriegel, F. Scotognella, L. Manna, Plasmonic doped semiconductor nanocrystals: Properties, fabrication, applications and perspectives, Physics Reports. 674 (2017) 1–52. https://doi.org/10.1016/j.physrep.2017.01.003.

[27] I. Kriegel, C. Urso, D. Viola, L. De Trizio, F. Scotognella, G. Cerullo, L. Manna, Ultrafast Photodoping and Plasmon Dynamics in Fluorine–Indium Codoped Cadmium Oxide Nanocrystals for All-Optical Signal Manipulation at Optical Communication Wavelengths, The Journal of Physical Chemistry Letters. 7 (2016) 3873–3881. https://doi.org/10.1021/acs.jpclett.6b01904.

[28] C. Schinke, P. Christian Peest, J. Schmidt, R. Brendel, K. Bothe, M.R. Vogt, I. Kröger, S. Winter, A. Schirmacher, S. Lim, H.T. Nguyen, D. MacDonald, Uncertainty analysis for the coefficient of band-to-band absorption of crystalline silicon, AIP Advances. 5 (2015) 067168. https://doi.org/10.1063/1.4923379.

[29] L.J. Phillips, A.M. Rashed, R.E. Treharne, J. Kay, P. Yates, I.Z. Mitrovic, A. Weerakkody, S. Hall, K. Durose, Dispersion relation data for methylammonium lead triiodide perovskite deposited on a (100) silicon wafer using a two-step vapour-phase reaction process, Data in Brief. 5 (2015) 926–928. https://doi.org/10.1016/j.dib.2015.10.026.

[30] L.J. Phillips, A.M. Rashed, R.E. Treharne, J. Kay, P. Yates, I.Z. Mitrovic, A. Weerakkody, S. Hall, K. Durose, Maximizing the optical performance of planar CH3NH3PbI3 hybrid perovskite heterojunction stacks, Solar Energy Materials and Solar Cells. 147 (2016) 327–333. https://doi.org/10.1016/j.solmat.2015.10.007.

[31] B. Park, Probing and passivating electron traps at the MAPbI3/TiO2 interface, Results in Physics. 23 (2021) 104025. https://doi.org/10.1016/j.rinp.2021.104025.

[32] T. Komikado, S. Yoshida, S. Umegaki, Surface-emitting distributed-feedback dye laser of a polymeric multilayer fabricated by spin coating, Appl. Phys. Lett. 89 (2006) 061123. https://doi.org/10.1063/1.2336740.